\documentclass{article}

\usepackage{arxiv}

\usepackage[utf8]{inputenc} %
\usepackage[T1]{fontenc}    %
\usepackage{hyperref}       %
\usepackage{url}            %
\usepackage{booktabs}       %
\usepackage{amsfonts}       %
\usepackage{nicefrac}       %
\usepackage{microtype}      %
\usepackage{lipsum}		%
\usepackage{graphicx}
\usepackage{natbib}
\usepackage{doi}
\usepackage{float}
\usepackage{tikz}
\usepackage{stfloats}
\usepackage{lipsum}
\usepackage{kantlipsum}
\usepackage{multicol}
\usepackage{graphicx,subcaption}
\usepackage{subcaption}
\usepackage{cleveref}
\usetikzlibrary{arrows,positioning,shapes.geometric}
\definecolor{lightblue}{RGB}{153,204,255}
\definecolor{lightgreen}{RGB}{53,255,153}

\title{An Algorithm for the Labeling and Interactive Visualization of the Cerebrovascular System of Ischemic Strokes}

\author{ \href{https://lme.tf.fau.de/person/thamm}{\hspace{1mm}Florian Thamm}\thanks{Corresponding Author;  \href{https://lme.tf.fau.de/person/thamm/}{https://lme.tf.fau.de/person/thamm/}} \\
	Pattern Recognition Lab\\
	FAU Erlangen-Nuremberg\\
	Erlangen, Germany\\
	\texttt{florian.thamm@fau.de} \\
	\And
	Markus J\"urgens \\
	Computed Tomography \\ 
    Siemens Healthcare GmbH \\ 
	Forchheim, Germany\\
	\And
	Oliver Taubmann \\
	Computed Tomography \\ 
    Siemens Healthcare GmbH \\ 
	Forchheim, Germany\\
	\And
	Aleksandra Thamm\\
	Pattern Recognition Lab\\
	FAU Erlangen-Nuremberg\\
	Erlangen, Germany\\
	\And
	Leonhard Rist\\
	Pattern Recognition Lab\\
	FAU Erlangen-Nuremberg\\
	Erlangen, Germany\\
	\And
  	Hendrik Ditt \\
	Computed Tomography \\ 
    Siemens Healthcare GmbH \\ 
	Forchheim, Germany\\
	\And
	Andreas Maier \\
	Pattern Recognition Lab \\
	FAU Erlangen-Nuremberg \\
	Erlangen, Germany\\
}

\date{}

\renewcommand{\shorttitle}{Labeling and Interactive Visualization of the Cerebral Vasculature}

\hypersetup{
pdftitle={Labeling and Interactive Visualization of the Cerebral Vasculature},
pdfsubject={eess.IV },
pdfauthor={Florian Thamm},
pdfkeywords={Computed Tomography Angiography, Ischemic Strokes, Segmentation, Labeling, Interactive Visualization},
}

\begin{document}
\maketitle

\begin{abstract}
During the diagnosis of ischemic strokes, the Circle of Willis and its surrounding vessels are the arteries of interest. Their visualization in case of an acute stroke is often enabled by Computed Tomography Angiography (CTA). Still, the identification and analysis of the cerebral arteries remain time consuming in such scans due to a large number of peripheral vessels which may disturb the visual impression. In previous work we proposed VirtualDSA++, an algorithm designed to segment and label the cerebrovascular tree on CTA scans. Especially with stroke patients, labeling is a delicate procedure, as in the worst case whole hemispheres may not be present due to impeded perfusion. Hence, we extended the labeling mechanism for the cerebral arteries to identify occluded vessels. In the work at hand, we place the algorithm in a clinical context by evaluating the labeling and occlusion detection on stroke patients, where we have achieved labeling sensitivities comparable to other works between 92\,\% and 95\,\%. To the best of our knowledge, ours is the first work to  address labeling and occlusion detection at once, whereby a sensitivity of 67\,\% and a specificity of 81\,\% were obtained for the latter. VirtualDSA++ also automatically segments and models the intracranial system, which we further used in a deep learning driven follow up work. We present the generic concept of iterative systematic search for pathways on all nodes of said model, which enables new interactive features. Exemplary, we derive in detail, firstly, the interactive planning of vascular interventions like the mechanical thrombectomy and secondly, the interactive suppression of vessel structures that are not of interest in diagnosing strokes (like veins). We discuss both features as well as further possibilities emerging from the proposed concept.
\end{abstract}


\section{Introduction}
If the cerebrovascular system takes damage or is disturbed, no blood and therefore no oxygen reaches the brain parenchyma causing brain cells to die. This medical condition is known as stroke and is further subdivided into two types. Strokes caused by spontaneous ruptures of the cerebral arteries often preceded by aneurysms are called the hemorrhagic type. The other type, the occlusion of arteries, which in this context refers to the complete blockage of a blood vessel caused by a thrombus or embolus, is called the ischemic type. Both types are usually diagnosed using Computed Tomography (CT). The hemorrhagic type can be detected on CT images without the use of a contrast agent (non-contrast CT images, short NCCT) with a sensitivity close to 100\,\% \citep{schellinger2007noninvasive}. Ischemic strokes usually require contrast-enhanced acquisitions, referred to as CT angiography (CTA) scans since the cerebrovascular system becomes visible. Its visualization serves two purposes. First, occlusions become visible as discontinuations on the vessel tree, enabling an accurate localization of the occlusion. Moreover, once the vessel tree becomes visible, mechanical thrombectomies, the physical removal of the clot, can be planned and the images used for guidance during the intervention.

However, a raw CTA scan does not provide a clear overview of the vessels without further processing. Bones, calcifications, and tissue with an attenuation comparable to that to contrast agent (e.g., iodine) disturb the visual impression significantly. Thus, it is common practice to subtract a CTA and NCCT scan digitally from each other, so only contrast-enhanced parts remain. These digitally subtracted angiography (DSA) scans come with some drawbacks such as additional radial exposure. Also, a DSA scan visualizes both veins and arteries, whereas the former are not always of interest and may disturb the visual impression during the diagnosis of the arterial branches.

In previous work, we addressed this issue by the proposal of a segmentation and labeling algorithm called VirtualDSA++ \citep{thamm}. As the labeling on ischemic stroke patients is delicate, we extended the labeling mechanism to consider occlusions. Once a large vessel is occluded, affected sub-trees visually disappear, which considerably changes the labeling situation. The work at hand brings the labeling and occlusion detection approach into a clinical context as we evaluated its performance on stroke patients. Next to the labeling, VirtualDSA++ allows users to interact with the vessel tree by planning pathways or reducing the intravascular tree to relevant sections for the stroke diagnosis, a feature we call vein suppression. In this work, we present the general concept of the iterative pathway search from which we derive exemplarily the path planning and vein suppression feature. Depending on the optimization criterion of the iterative search, further applications are possible which will be discussed as well. Finally, we will provide a much more comprehensive overview as well as more detailed descriptions compared to the brief introduction of VirtualDSA++ given in \cite{thamm}.

In a follow up work, VirtualDSA++ was one of the essential steps for a deep learning based domain transfer task, published in \cite{thamm2021syncct}. We synthesised NCCT images from CTA scans, by removing all contrast enhanced vessel structures, enforced by a GAN-learning mechanism, while using just one single energy level. This was only possible by the prior segmentation of the vessel tree through VirtualDSA++, as it served as guidance for structured to be removed during the transfer.

\section{Related Works}
\subsection{Segmentation}
The vast majority of the works targeting the segmentation of the intracranial system are based on MRI angiography, to be more specific on time-of-flight (TOF) acquisitions \citep{Livne.2019, Li.2017, Giles., Hilbert.2020, Chen.}. During a TOF scan, stationary tissue stays saturated by the radio frequency pulses in contrast to flowing liquids which do not experience any excitation by the same pulses. Moving liquids thus keep the high initial magnetization as they flow through excited slices. As a result, blood appears significantly brighter than the surrounding tissue. The advantage of such a scan is that no tissue other than flowing blood significantly interferes with the visual appearance of the angiographic scan -- ideal for segmentation tasks. Contrary to CTA scans, where contrast agent briefly increases the average attenuation of blood, allowing it to be distinguished from brain parenchyma but not (without considering anatomy, based only on HU values) from dense structures like bones. This renders the automated segmentation of the vessel tree on CTA scans substantially more difficult. Nevertheless, CTA is commonly used to diagnose strokes as it is fast, highly available, and comparably cheap to maintain. In many clinics, an NCCT scan is performed first to determine the underlying stroke type, followed by a CTA scan if necessary. Works solely utilizing CTA scans are quite rare, as often additional NCCT scans are incorporated into the pipeline as well, like in \cite{Manniesing.2005, Hernandez.2007}. Despite the difficulty to segment the vessel tree on CTA data alone, an early feasibility study \citep{feasability} has shown a segmentation of the cerebral arterial tree with no additional NCCT is possible. As a successor \cite{Nazir.2020} proposed a patch-wise segmentation network similar to a ResNet consisting of Inception modules \citep{ChristianSzegedy.2015}. In a cross-validation on 70 data sets their approach achieved a Dice score of 89.46\%. Still, the development of supervised segmentation networks is compounded by the fact that the labeling itself requires an extraordinary effort, as \cite{Nazir.2020} reported. The labeling of one data set took 6 hours, resulting in roughly a month of work required to annotate 20 cases \citep{Nazir.2020}.

\subsection{Labeling}
MRI data has not only been used for the segmentation but consequently for the labeling too. \cite{vandeGiessen.} as well as \cite{robben2013anatomical, bogunovic2013anatomical} worked with TOF-MRI data while \cite{Dunas.2016, Dunas.2017} worked with 4D-flow MRI data. Both works were made possible by using an atlas which serves as guidance for the classification of each vessel section, achieving sensitivities between 90\% and 100\%. Likewise, with the segmentation of the intracranial system, MRI has the advantage that no tissue other than blood interferes with the segmentation significantly. \cite{shen2020automatic} used key points combined with a path finding algorithm in order to identify intracranial vessels in CTA scans for the purpose of path disentanglement. Nevertheless, \cite{shen2020automatic}'s approach does not cover the segmentation itself and assumes it is given. 

Notably, none of the prior work we are aware of evaluates their approach on stroke cases, which is crucial for two key reasons. Firstly, since the occlusion present in an ischemic stroke causes parts of a patient's vessel tree to ``disappear'' due to lack of contrast, rendering the problem substantially more difficult. Secondly, analysis of the vessel tree is of particular interest specifically in stroke patients for comprehensive diagnosis and ideal treatment planning.

\subsection{LVO Detection}
Research has been done in the detection of LVOs as well. \cite{amukotuwa2019fast} presented a pipeline tailored explicitly for the detection of LVOs and the affected brain region. The authors conducted two studies with two different data cohorts, for which they achieved a ROC-AUC of 0.86\,\% on 477 patients in the first \citep{amukotuwa2019automated} and a ROC-AUC of 0.94\,\% on 926 patients \citep{amukotuwa2019fast} in the second study. While \cite{amukotuwa2019automated} relied solely on image processing, deep learning methods have also been developed. For instance, \cite{stib2020detecting} used multi-phase CTA scans in order to compute maximum-intensity projections of segmentations to finally classify the presence of LVOs with 2D-DenseNets receiving one or more of the different CTA phases (arterial, peak, and late venous). The ROC-AUC they achieved ranges from 0.74\,\% (worst phase constellation using late venous) to 0.85\,\% (best phase using the peak venous phase) on 424 patients. In \cite{Sheth.2019} a convolutional neural network (CNN) processes both hemispheres separately with shared kernel weights before a classification head predicts the presence of LVOs achieving an AUC of 0.88\,\%.

\subsection{Pathway Search}
Path finding algorithms are commonly used in various kinds of applications. On cerebrovascular vessel trees, \cite{Suran.} performed path search based on MRI angiography data. They considered the binary skeleton of a vessel tree as a graph, where each white pixel represents one node of the graph. In order to determine the shortest pathways, the Dijkstra algorithm is applied to the pixel-graph. \cite{shen2020automatic} used Dijkstra as well for their vessel straightening approach, applying it to a sub-sampled skeleton of the vessel tree. Computing paths on a pixel basis and even on sub-sampled skeletons is inefficient as the number of operations of the search algorithm increases with the number of nodes. Also, the Dijkstra algorithm is not the best choice for this particular problem as we will discuss in Sec.~\ref{discussion}. In Sec.~\ref{SearchOnModels}, the work at hand therefore presents two improvements to boost the efficiency and run time, allowing to compute paths interactively.
\section{Data}
\label{Data}
In total, 171 CTA scans covering the entire head region, acquired at a single clinical site, were available, distributed over 83 male and 88 female patients with an age of $71\pm 13$ years ranging from 27 to 96 years. In 15 cases, contrast agent injection was unsuccessful, strong metal artifacts disturbed critical regions, or the registration to the atlas (see Sec.~\ref{SegmentationPipeline}) failed. Of the remaining 156, 99 patients (57\%) were LVO positive and had occlusions in the middle cerebral artery (MCA M1 or proximal M2) or the internal carotid artery (ICA). One patient had a complete occlusion in the Basilar Artery with a consecutive under-perfusion in the left and right posterior cerebral artery (PCAL and PCAR). No subjects had an occlusion in the Anterior Cerebral Artery (ACA). All data was acquired using a Somatom Definiton AS+ (Siemens Healthineers, Forchheim, Germany) and reconstructed with a slice thickness of either 1\,mm (15 data sets) or 0.5\,mm (141 data sets). Both are considered thin slice reconstructions that ensure an accurate distinction of dense nearby vessel branches. %
\section{Methods}
The pipeline of VirtualDSA++, shown in Fig.~\ref{pipe}, can be subdivided into three stages and is described in section \ref{SegmentationPipeline}. The segmentation of the cerebrovascular vessel tree is enabled by region growing on a bone-free CTA scan as indicated by the ``Segmentation'' stage in Fig.~\ref{pipe}. The seed points for the final region growing are automatically determined in the ``Seed Points'' stage of the pipeline, which overall performs the majority of processing steps. Intermediate results of the seed point generation are further used to enable the automated labeling of the large vessels close to the Circle of Willis, presented in Sec.~\ref{VesselLabeling}. The surface and the skeleton of the determined segmentation are computed in a next step using MeVisLab which allows further algorithmic processing in a third stage called ``Modelling'', where path search algorithms as well as the selective vein masking can be applied in an interactive setup. Sec.~\ref{Algorithms} presents the proposed approach to enable these features using the A* path search algorithm.
\subsection{Segmentation}
\label{SegmentationPipeline}
\begin{figure} 
  \label{ fig7}
    \begin{minipage}[b]{0.54\linewidth}
    \centering

\begin{tikzpicture}
\tikzset{block/.style= {draw, rectangle, thick, align=center, minimum width=1cm, minimum height=0.5cm},}
    \node[block, fill = lightgray]                               (cta) {CTA};
    \node[block, fill = lightgray, right = 1cm of cta]           (brainAtlas) {Brain  Atlas};
    \node[block, fill = lightgray ,right = 1cm of brainAtlas]    (vesselAtlas) {Vessel  Atlas};
    
    \node[block, below = 1cm of cta]             (br) {Bone Removal$_{(a)}$};
    \node[block, below = 1cm of brainAtlas]      (reg) {Reg$_{(b)}$};
    \node[block, below = 1cm of vesselAtlas]     (trafo) {Trafo$_{(c)}$};
    
    \node[block, below = 1cm of reg]    (frangi){Frangi$_{(d)}$};
    \node[block, below = 1cm of trafo] (t1){$> t_1$};
    
    \node[block, below = 0.5cm of t1] (dilat){Dilat};
    \node[block] at (dilat -| reg) (mask1) {Mask$_{(e)}$};

    \node[block, below = 0.5cm of mask1] (t2) {$> t_2$};
    
    \node[coordinate, below = 1cm of t2] (t2_hough){};
    \node[block] at (t2_hough -| dilat) (hough){Hough$_{(f)}$};
    
    \node[block, below = 0.5cm of hough] (mask2){Mask$_{(g)}$};
    
    \node[coordinate, below = 0.5cm of mask2] (mask2_rg){};
    \node[block] at (mask2_rg -| t2) (rg1){RG$_{(h)}$};
    
    \node[coordinate, below = 0.5cm of rg1](rg1_rg2){};
    \node[block] at (br |- rg1_rg2) (rg2) {RG$_{(i)}$};
    
    \node[block, below = 1cm of rg2](closing){Closing};
    \node[block] at (closing -| mask2) (labeling) {Labeling};
    
    \node[block, below = 0.5cm of closing](dtf){Dtf$_{(j)}$};
    \node[block, fill = lightgray, below = 0.5cm of labeling](labelingo){Labels};
    
    \node[block, below = 0.5cm of dtf](graph){Graph\\ Search$_{(k)}$};
    \node[block, fill = lightgray] at (dtf -| rg1) (segment) {Segmentation};

    \node[block, fill = lightgray] at (graph -| segment) (modell) {Model};

    \node[coordinate, below = 0.5cm of cta](cta_b){};
    \node[coordinate, below = 0.5cm of brainAtlas](brainAtlas_b){};
    \node[coordinate, left  = 0.1cm of brainAtlas_b](brainAtlas_b_l){};
    \node[coordinate, below  = 0.5cm of brainAtlas_b_l](brainAtlas_b_l2){};
    
    \node[coordinate, right = 0.1cm of brainAtlas_b](brainAtlas_b_r){};
    \node[coordinate, above = 0.5cm of brainAtlas_b_r](brainAtlas_b_r1){};
    \node[coordinate, below = 1cm of brainAtlas_b_r1](brainAtlas_b_r2){};
    
    \node[coordinate, below = 0.5cm of frangi](frangi_b){};
    
    \node[coordinate, below = 0.5cm of t2](t2_b){};
    
    \node[coordinate, below = 0.5cm of br](br_b){};
    \node[coordinate, left = 1cm of br_b](br_b_l){};
    
    \node[coordinate, left = 1cm of rg2] (rg2_l){};
    
    \node[coordinate, below = 0.5cm of reg](reg_b){};
    \node[coordinate, below = 0.5cm of rg1](rg1_b){};
    \node[coordinate, below = 0.5cm of rg2](rg2_b){};

     \path[draw, ->]
        (cta) edge (br)
        (cta_b) -- (brainAtlas_b_l)
        (brainAtlas_b_l) edge (brainAtlas_b_l2)
        (brainAtlas_b_r1) edge (brainAtlas_b_r2)
        (vesselAtlas) edge (trafo)
        (reg) edge (trafo)
        (trafo) edge (t1)
        (t1) edge (dilat)
        (dilat) edge (mask1)
        (br) -- (br_b)
        (br_b) -- (reg_b)
        (reg_b) edge (frangi)
        (frangi) edge (mask1)
        (mask1) edge (t2)
        (t2) edge (rg1)
        (t2_b) -- (t2_b -| hough)
        (t2_b -| hough) edge (hough)
        (hough) edge (mask2)
        (rg1 |- mask2) edge (mask2)
        (mask2) -- (mask2 |- rg1)
        (mask2 |- rg1) edge (rg1)
        (rg1) -- (rg1_b)
        (rg1_b |- rg2) edge (rg2)
        (rg1_b) -- (rg1_b -| mask2)
        (rg1_b -| mask2) edge (labeling)
        (rg2) edge (closing)
        (rg2_b) -- (rg2_b -| rg1)
        (rg2_b -| rg1) edge (segment)
        (closing) edge (dtf)
        (dtf) edge (graph)
        (graph) edge (modell)
        (labeling) edge (labelingo)
        (br_b) -- (br_b |- rg1)
        (rg2 |- rg1) -- (rg2)
    ;
    
    \filldraw[draw = blue, fill=blue, very thick, fill opacity = 0.1, draw opacity = 0.1](1.6,-1.1) rectangle (-1.6, -10.2);
    \filldraw[draw = orange, fill=orange, very thick, fill opacity = 0.1, draw opacity = 0.1](7,-1.1) rectangle (1.7, -10.2);
    \filldraw[draw = green, fill=green, very thick, fill opacity = 0.1, draw opacity = 0.1](1.3,-10.8) rectangle (-1.6, -14.8);
    \node[color=gray] at (4.6, -5.2) {Seed Points};
    \node[color=gray, rotate = 90] at (-1.3, -5) {Segmentation};
    \node[color=gray, rotate = 90] at (-1.3, -12.5) {Modelling};
    
\end{tikzpicture}
\caption{The proposed processing pipeline of VirtualDSA++, showing the three sections: the segmentation, enabled by the seed point finding and the modelling of the vessel tree at the end. Each box represent one step of the pipeline. Image inputs of each step are represented by incoming arrows from the top. Secondary information, like seed points for region growing (abbreviated as RG) or the transformation matrix step (c), enter the box horizontally.}

\label{pipe}

    \vspace{4ex}
  \end{minipage}
  \begin{minipage}[b]{0.44\linewidth}
    \centering

\begin{subfigure}[t]{.42\linewidth}
\centering
\includegraphics[width=\textwidth]{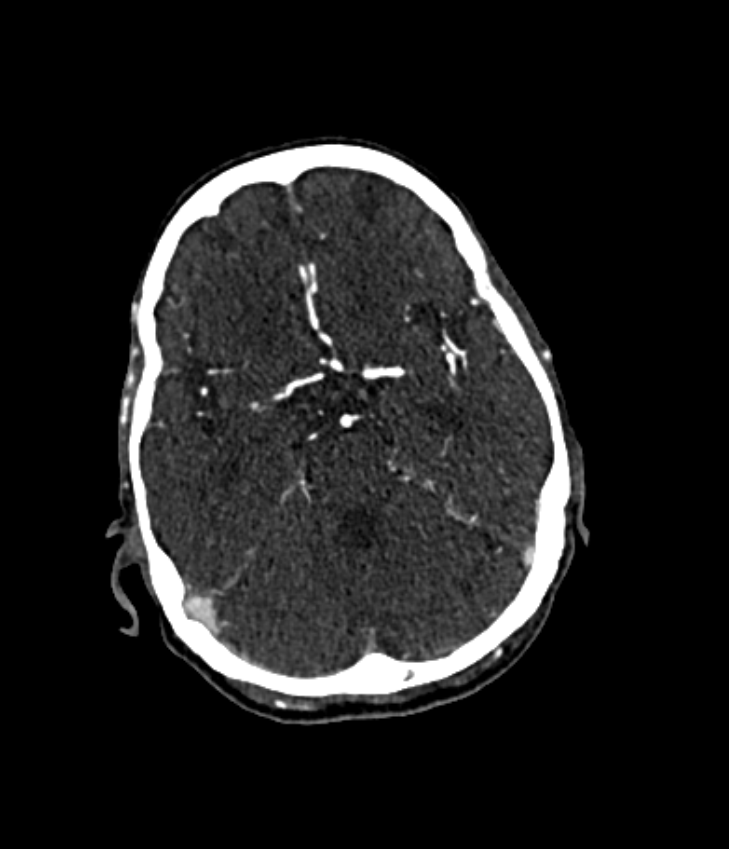}
\caption{CTA}
\label{fig:cta}
\end{subfigure}
\begin{subfigure}[t]{.42\linewidth}
\centering
\includegraphics[width=\textwidth]{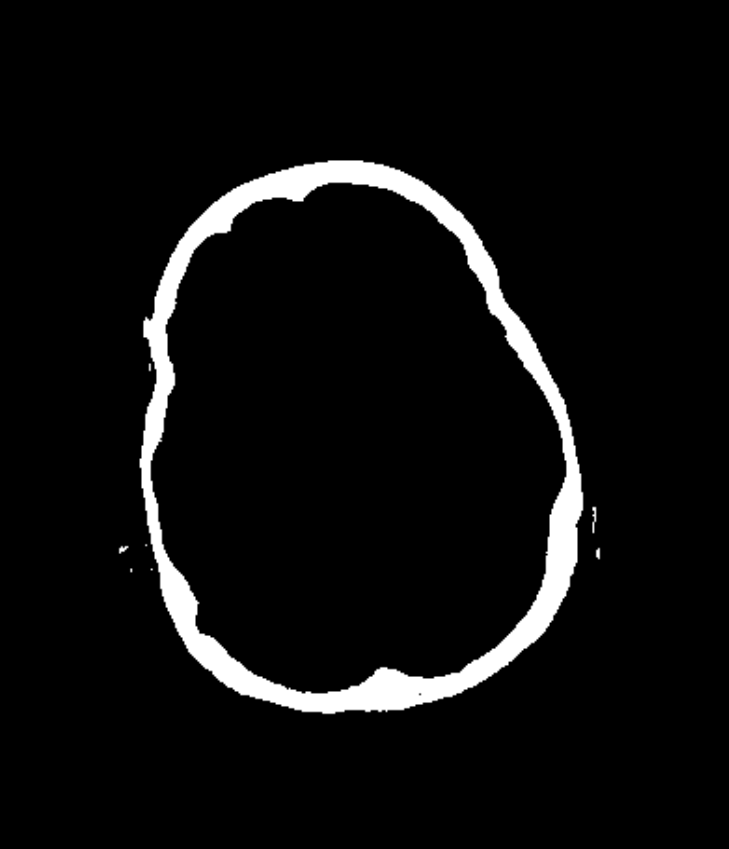}
\caption{Bone Removal Mask $_{(a)}$}
\label{fig:br}
\end{subfigure}
\begin{subfigure}[t]{.42\linewidth}
\centering
\includegraphics[width=\textwidth]{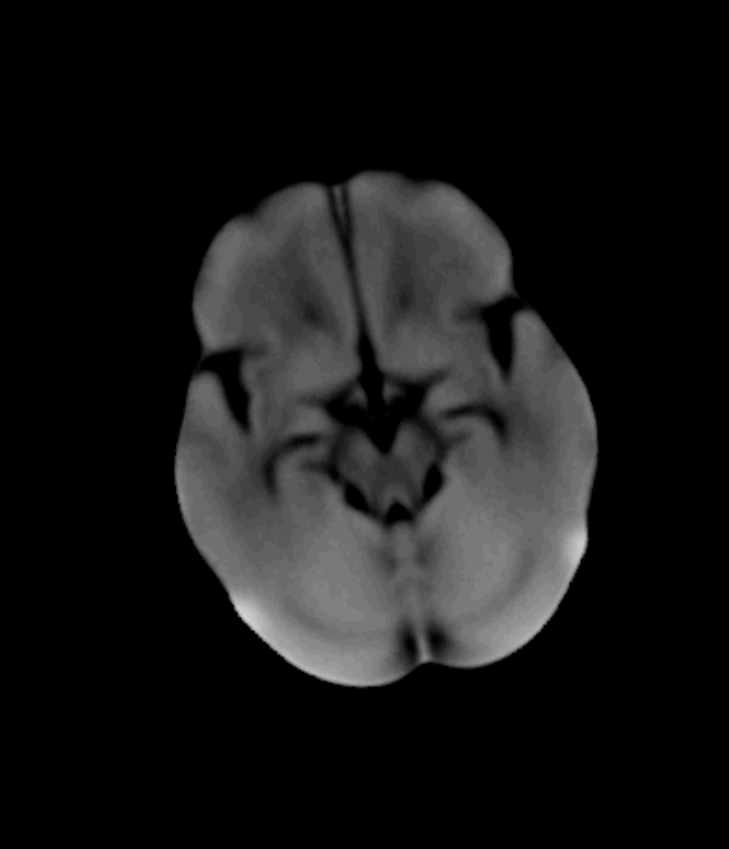}
\caption{Registered Brain Atlas}
\label{fig:brain}
\end{subfigure}
\begin{subfigure}[t]{.42\linewidth}
\centering
\includegraphics[width=\textwidth]{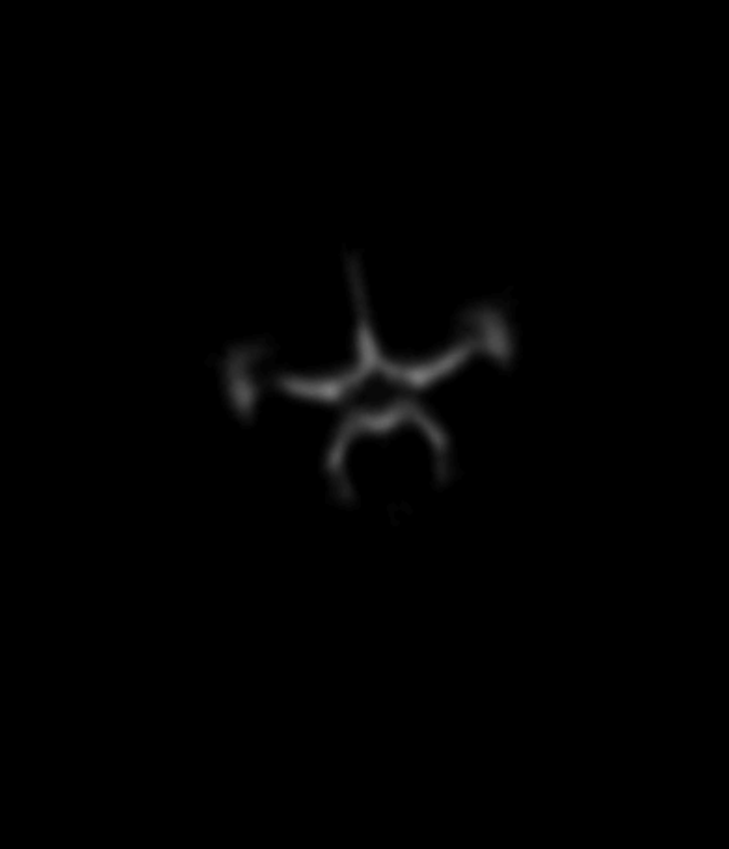}
\caption{Transf. Vessel Atlas}
\label{fig:vessel}
\end{subfigure}
\begin{subfigure}[b]{.42\linewidth}
\centering
\includegraphics[width=\textwidth]{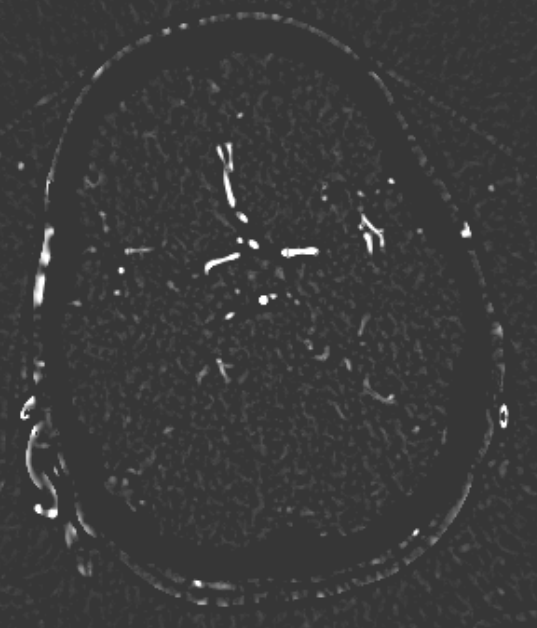}
\caption{Frangi$_{d}$}
\label{fig:frangi}
\end{subfigure}
\begin{subfigure}[b]{.42\linewidth}
\centering
\includegraphics[width=\textwidth]{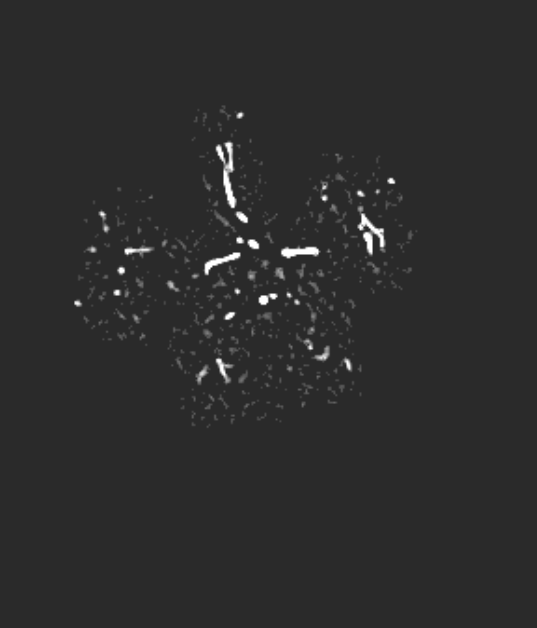}
\caption{Mask$_{(e)}$}
\label{fig:mask_e}
\end{subfigure}
\begin{subfigure}{.42\linewidth}
\centering
\includegraphics[width=\textwidth]{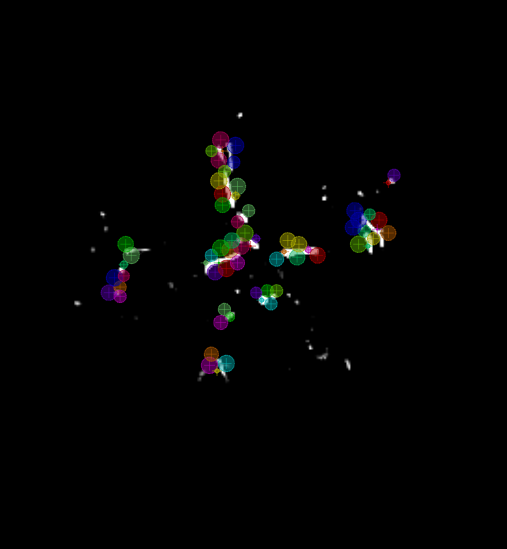}
\caption{Hough$_{(f)}$}
\label{fig:hough}
\end{subfigure}
\begin{subfigure}{.42\linewidth}
\centering
\includegraphics[width=\textwidth]{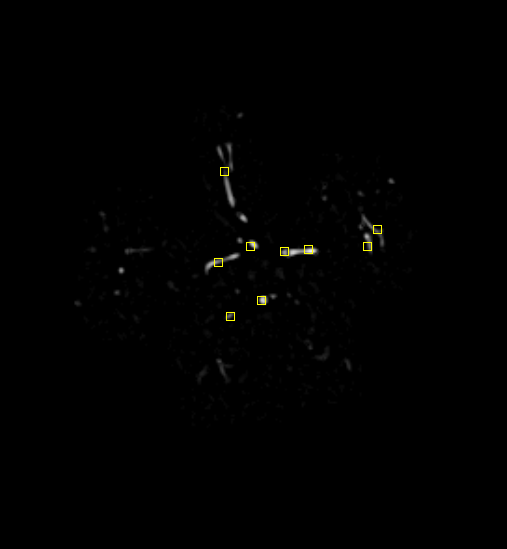}
\caption{Mask$_{(g)}$}
\label{fig:mask_g}
\end{subfigure}
\begin{subfigure}{.42\linewidth}
\centering
\includegraphics[width=\textwidth]{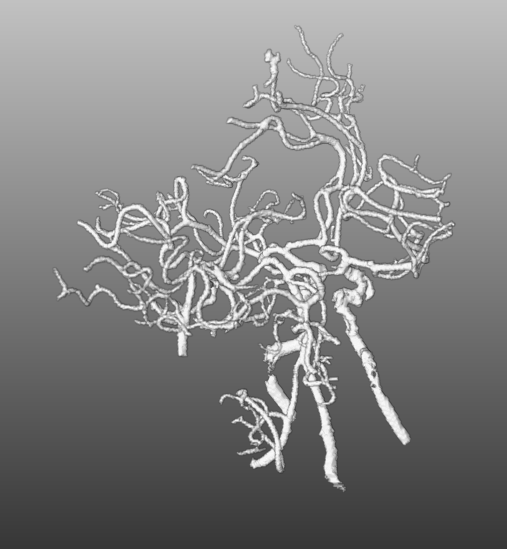}
\caption{RG$_{(h)}$}
\label{fig:rg_h}
\end{subfigure}
\begin{subfigure}{.42\linewidth}
\centering
\includegraphics[width=\textwidth]{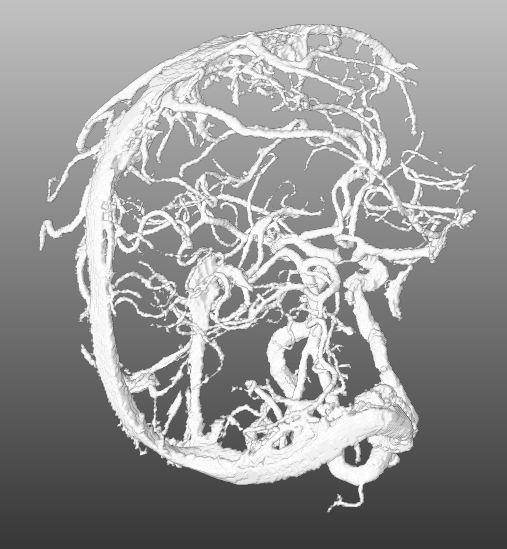}
\caption{RG$_{(i)}$}
\label{fig:rg_i}
\end{subfigure}
\caption{Inputs and intermediate results of the VirtualDSA++ pipeline, which is described in Fig.~\ref{pipe}.}
  \end{minipage}%

\end{figure}
  
The pipeline we propose receives a CTA image as input, an example of which is shown in Fig.~\ref{fig:cta}. Alongside the CTA data, a brain atlas (shown already registered to an example in Fig.~\ref{fig:brain}, a mean brain over a patient populations) \citep{Kemmling.2012} together with the corresponding vessel atlas \citep{Mantas.2012} (accordingly transformed, shown in Fig.~\ref{fig:vessel}) is given. The latter atlas is a volume representing the existence of voxels belonging to a vessel by means of a probability value, whereby it emphasizes on the arterial branches close to the Circle of Willis. In a first step, all bone structures are removed, as indicated in Fig.~\ref{pipe}(a), using a method by \cite{mingqing.2020} which segments those bone structures in CTA scans. An example is shown in Fig.~\ref{fig:br} which is the bone segmentation of the slice shown in Fig.~\ref{fig:cta}. The CTA volume is not only forwarded to the bone removal but also used to register the brain atlas onto in Fig.~\ref{pipe}(b). The non-rigid (elastic) registration is based on \cite{chefd2002flows}'s method specifically designed to register multi-modal volumes. The resulting deformation field is used to transform the vessel atlas \citep{Mantas.2012} in step Fig.~\ref{pipe}(c) into the CTA coordinate system. Tubular structures of the bone-free CTA scan,  resulting from Fig.~\ref{pipe}(a), are enhanced using the Frangi filter \citep{frangi1998multiscale} (two scales, $\sigma_1 = 1.00$, $\sigma_u = 1.50$) exemplarily shown in Fig.~\ref{fig:frangi}. The filter response is masked in Fig.~\ref{pipe}(e) with the vessel atlas, which is binarized with a relative threshold of $t_1 = 0.5\%$ (of the maximum value) followed by dilation (with a kernel size in z,y and x of 11, 7 and 7, respectively). Fig.~\ref{fig:mask_e} shows the result of the vessel masked Frangi response, which is thresholded with $t_2 = 4$. Values below that threshold are suppressed to zero, values above are kept with the original values. With this step, noise is being reduced which stabilizes the subsequent steps, like the slice-wise circle based Hough transformation in step Fig.~\ref{pipe}(f)(canny threshold = 10, accumulator threshold = 1, min distance = 5, min radius = 0, max radius = 5 and accumulator threshold = 1). The goal of the Hough transformation is to identify circle-shaped structures in the given slices, however, as shown in Fig.~\ref{fig:hough} many centerpoints of the detected circles do not match with the preliminary vessel tree segmentation. Hence in the next step, the mask used for the Hough transform is reused again, to mask out in step Fig.~\ref{pipe}(g) all centerpoints which do not align with the segmentation. The example shown in Fig.~\ref{fig:hough} originally has 6747 centerpoints in total, reduced to 1375 by the masking, the result of which is shown in Fig.~\ref{fig:mask_g}. The remaining points are used as seed points for region growing in Fig.~\ref{pipe}(h). The region growing segments all voxels connected to the seed points, whose intensity differs no more than $\pm 5\%$ from the average intensity of the seed point voxel values. The result is a preliminary segmentation which only includes vessel segments represented in the original vessel atlas. As will be described below, this mask is used for labeling. The result is shown in Fig.~\ref{fig:rg_h}. In order to segment all vessels, especially more distant vessel structures, the segmentation mask of step Fig.~\ref{pipe}(h) is transformed into seed points which are used for the final region growing in Fig.~\ref{pipe}(h) segmenting every connected voxel with a value in the range of 130 HU to 1500 HU. Its result is the final segmentation, shown in Fig.~\ref{fig:rg_i}. The whole pipeline has been implemented in MeVisLab 3.4, MeVis Medical Solutions AG, Bremen. 

The remaining steps, modelling and labeling, are described in Sec.~\ref{Algorithms} and  Sec.~\ref{VesselLabeling}, respectively.

\subsection{Modelling}
\label{Algorithms}
\begin{figure*}[t]
    \centering
    \includegraphics[width = \textwidth]{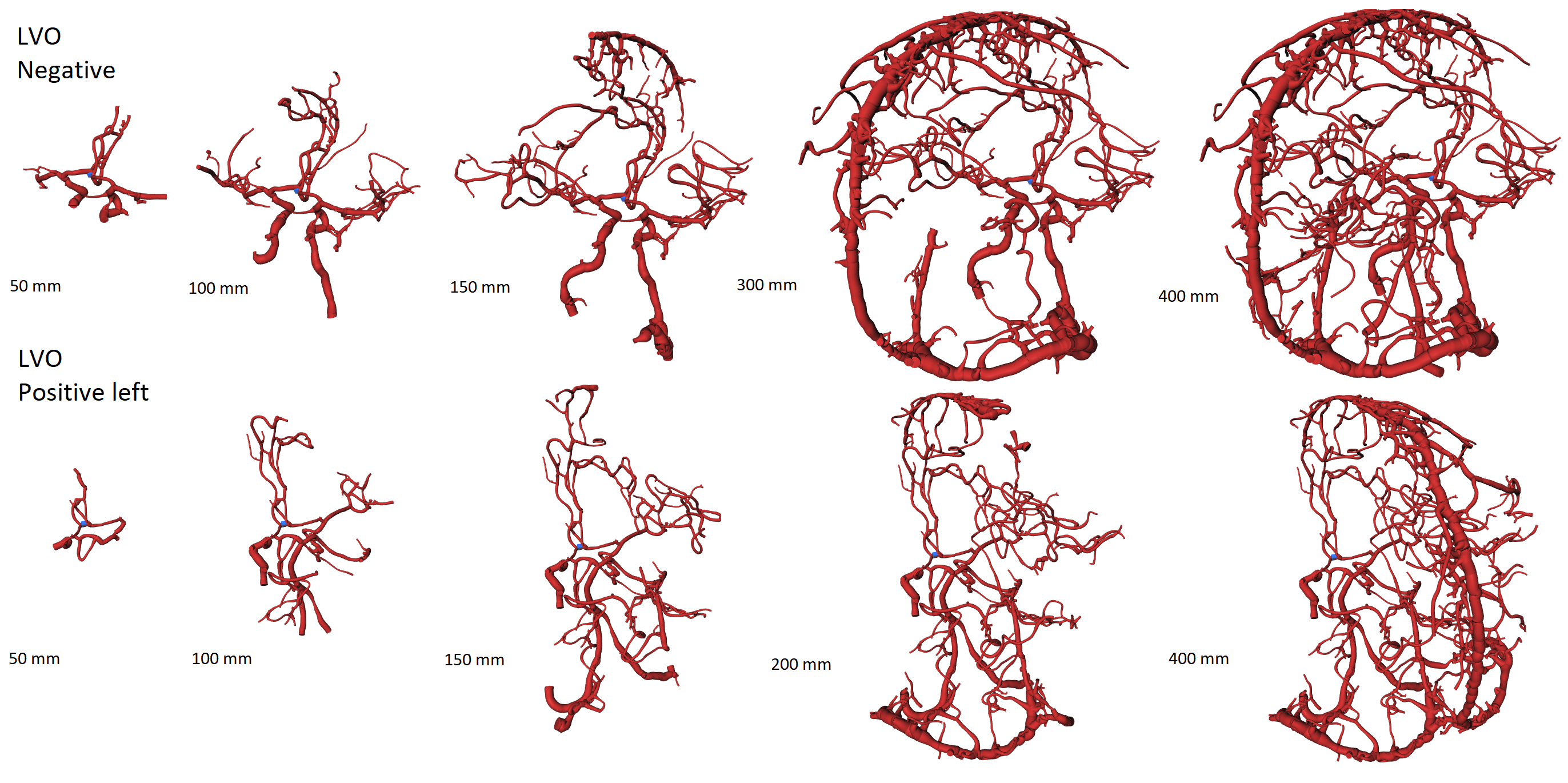}
    \caption{The surface model of the vessel tree of two patients demonstrating the vein suppression feature on different rendering distances. The upper row is an example of a patient with no LVO present. Contrary to the lower row, where the patient suffers an LVO on the left MCA branch. The expansion on the far right shows the full vessel tree including all veins. Relevant structures close to the Circle of Willis are visually obstructed. 
    }
    \label{fig:VeinSuppression}
\end{figure*}

For the interactive part of our work, the segmentation must be modeles as geometric representation e.g. as graph of centerlines and lumen radii. First, as tiny gaps may occur in the segmentation, closing is applied, followed by the Distance-Transform-Skeletonization algorithm (short, DTF-Skeletonization) \citep{DirkSelle.2000}, which returns not only the skeleton itself but also a surface model of the vascular structures. Examples trees are shown in Fig.~\ref{fig:VeinSuppression}, Fig.~\ref{MarkerResult}, Fig.~\ref{MultiMarker} and Fig.~\ref{fig:distanceIncrease}(a). Once the model is available, the information required of the interactive masking (geodesic distances) is computed in Fig.~\ref{pipe}(k). As the processing consists primarily of searching pathways on the skeleton representation, the step is called ``Graph Search''. In Sec.~\ref{SearchOnModels} we present the generic concept of pathway searches on vessel trees, from which it is possible to derive applications. Exemplarily, we present two applications consequently following from the formal concept introduction, namely the interactive path search in Sec.~\ref{PathSearch} and the interactive suppression of veins in Sec.~\ref{VeinSuppression}.

\subsubsection{Search on Models}
\label{SearchOnModels}

The computed model represented as the skeleton and the surface of the cerebrovascular segmentation serves as a data representation from which new applications emerge. The skeleton is now referred to as a graph in the following, whereby bifurcations represent nodes and vessel segments between two bifurcations are considered edges. Further applications are enabled by the systematic search for optimal pathways between two or more nodes. A path is considered to be optimal if said path fulfills an optimality criterion. Here, the shortest pathway between two nodes in the vessel graph is the objective of optimization. The search is enabled by using the A* search algorithm with the Euclidean distance as heuristic meeting all optimality conditions to achieve optimal pathways while being computationally efficient at the same time. 

We determine the shortest paths from a single root node to all other nodes in the graph. If the iterative path search from the root node to all other nodes is terminated, the following information is available for each node independent of the actual optimality criterion:
\begin{enumerate}
    \item The sequence of nodes describing the path to the root node according to the optimization criterion. Here, the shortest path between one node and the root node. 

    \item The optimality value, e.g.~the path length of the shortest path found. 

    \item The direction of each edge attached to the node according to the optimality criterion. 

\end{enumerate}

Once the search has been done for all nodes, the three properties are cached for each node to provide quick access during interactions with the tree. In some cases, trees in the volume are not topologically attached to the tree where the root node belongs. Here, depending on the optimality criterion, applying the above algorithm by defining quasi-optimal nodes in those unattached trees is possible. In the case of the shortest-pathway criterion, a quasi-optimal node would be the closest possible node in the unattached tree to the main graph. Finally, these quasi-optimal nodes are treated as root nodes within their tree, so the iterative search can also be applied here. However, on unattached trees, the properties hold only for the new quasi-optimal root node.

\subsubsection{Vein Suppression}
\label{VeinSuppression}
In some cases, for instance if large contrast boluses were injected, veins like the sinus sagittalis may disturb the visual impression. The second search node property enables an easy way of displaying only relevant parts by placing the root node into the Circle of Willis and visualizing only nodes and their paths accessible within a certain geodesic distance. Fig.~\ref{fig:VeinSuppression} shows two trees on different geodesic distances. The lower tree is from a patient who suffers an LVO on the left MCA branch, while the top one is LVO negative, showing no noticeable defects. As displayed, veins and more distant vessels may conceal stroke-relevant parts in their full representation on the right end and disappear towards the left side while the distance is decreased.

\begin{figure}[t] 
 \begin{minipage}{\linewidth}
      \centering
      \begin{minipage}[t]{.45\textwidth}

            \includegraphics[width =.95\textwidth]{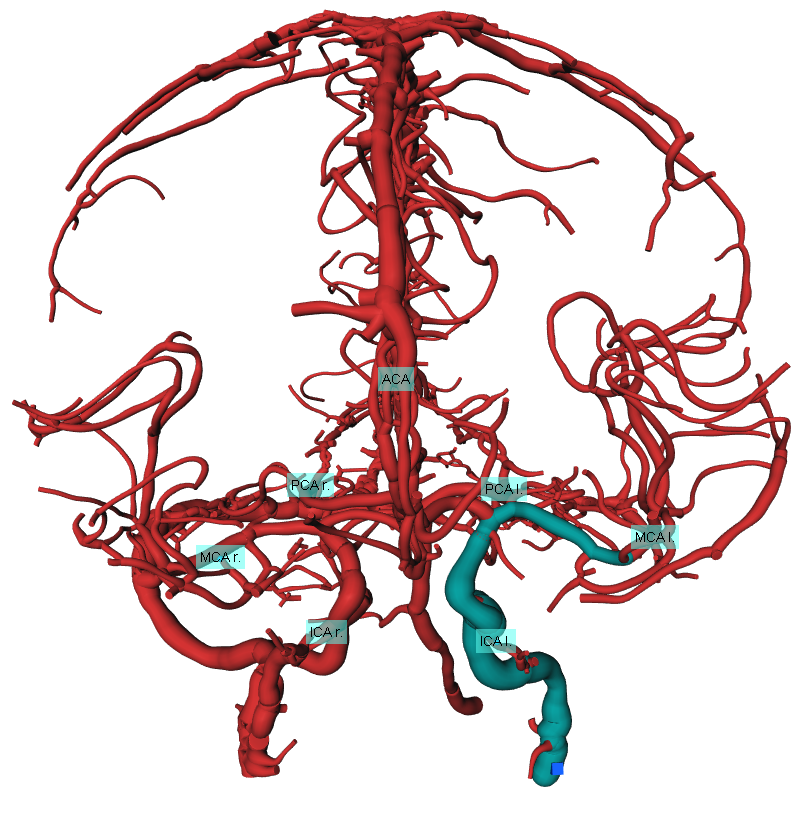}
	\caption{VirtualDSA++ model showing the ICA, MCA and ACA together with the placed markers. Highlighted in green is a path computed by the path search algorithm.}
	\label{MarkerResult}

      \end{minipage}
      \hspace{0.05\linewidth}
    \begin{minipage}[t]{.45\textwidth}

    \centering
    \includegraphics[width=.95\linewidth]{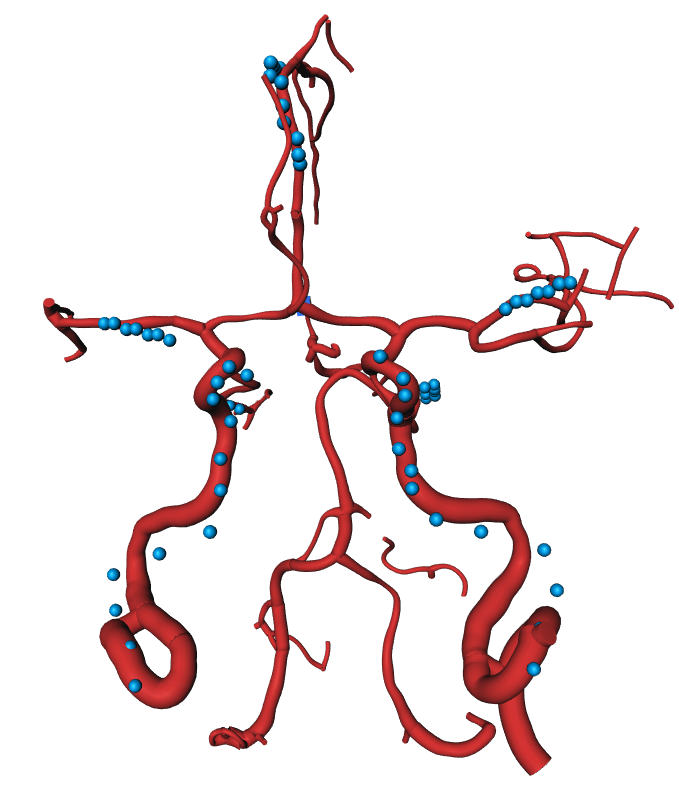}
   \caption{VirtualDSA++ model of the Circle of Willis, the ICA, MCA and ACA together with their transformed marker chains.}
	\label{MultiMarker}

      \end{minipage}
  \end{minipage}
\end{figure}

\subsubsection{Path Search}
\label{PathSearch}
Proper path planning is crucial when planning mechanical thrombectomy, i.e.~the physical removal of the blood clot causing the ischemia. Often, the catheter enters the intracranial system via the ICA to remove clots in it or in the MCA. The latter case is shown in Fig.~\ref{PathSearch}, where the root marker is displayed as a blue cube on the bottom right. In green the shortest path from the ICA root to the MCA is visualized. As the path search has been done and all properties are determined for all nodes, arbitrary shortest paths with respect to the root node can be visualized interactively in real-time. The path search can be used to readily annotate vessel segments for the purpose of documenting the findings, e.g.~for reporting.

\subsection{Vessel Labeling}

\label{VesselLabeling}

With the use of a cerebrovascular atlas, labeling can be done as well. With specific rules described in this section, it is further possible to detect missing or occluded vessels as well. That allows identifying patients suffering from LVOs that become visible as discontinuations in the vessel pathway. The idea is to place markers in the atlas space for each vessel that needs to be labeled and transform them into the desired CTA volume. Fig.~\ref{MultiMarker} shows the markers on the ICA left and right (bottom vertical paths), MCA left and right (transversal center pathways), and ACA (top vertical branch). 

To determine the presence or absence of a vessel, we propose the following approach. Each marker is assigned a fixed per-vessel maximum allowed distance. For each marker, the shortest distance to the vessel mask is computed. If the mask is within each marker's maximum allowed distance range for a certain number of markers per vessel, that vessel is identified as present. If not enough markers are close enough to the mask, the vessel is considered as missing or occluded. The positions, distances, and the number of markers per vessels are available upon request. To place the final labeling marker for visualization purposes, either the marker with the shortest distance can be selected or the voxel on the segmentation mask with the closest distance to the marker set. Final labeling markers are shown in Fig.~\ref{MarkerResult} for MCA l. and r., PCA l. and r., ICA l. and r. and ACA viewed sagital posterior on the vessel tree model.

\begin{figure*}[tb]
  \begin{subfigure}[b]{.49\linewidth}
    \centering\includegraphics[width=.95\linewidth]{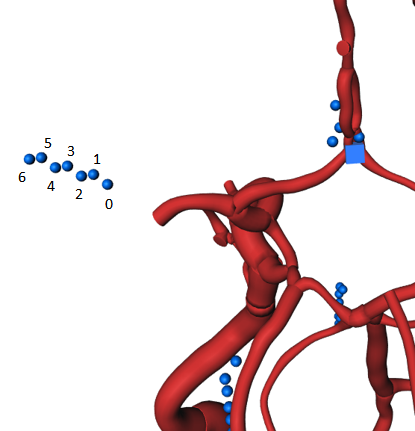}
    \caption{VirtualDSA++ segmentation of an occluded Middle Cerebral Artery with the corresponding Multi Marker.}
    \label{fig:markerIdx}
  \end{subfigure}
  \begin{subfigure}[b]{.49\linewidth}
    \centering\includegraphics[width=.95\linewidth]{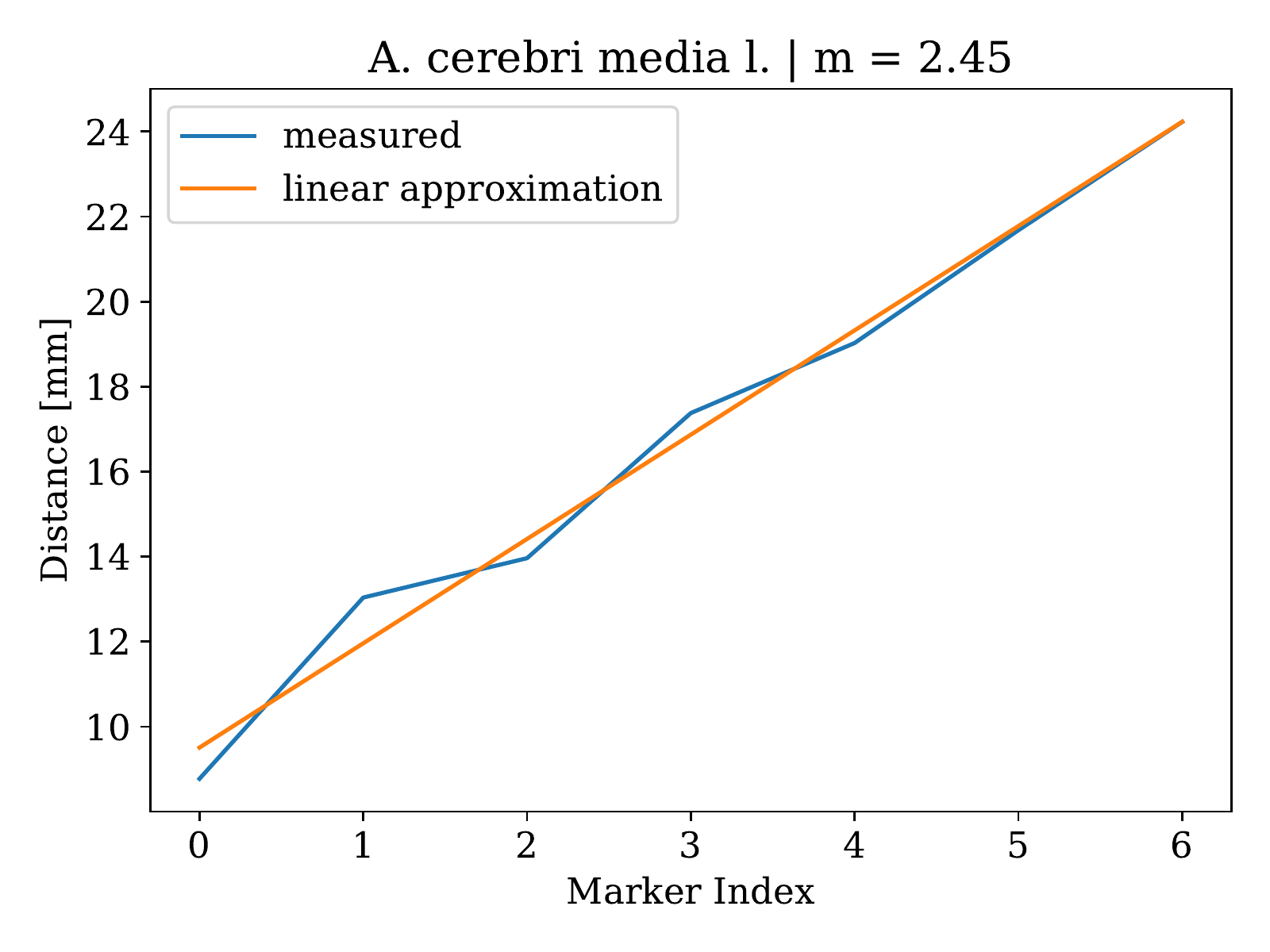}
    \caption{Relation between the ascending marker index with their distances to the closest vessel wall of the Middle Cerebral Artery shown in (a). The slope is 2.45 which means the distance increases with the marker index, thus an interruption in the vessel flow and consequently an occlusion is to be expected.}
    \label{fig:markerDis}
  \end{subfigure}
 \caption{Relation between the ascending marker indices and their distance to the closest vessel wall in case of large vessel occlusions, here the left Middle Cerebral Artery}
\label{fig:distanceIncrease}
\end{figure*}

We defined an additional criterion for the MCA (M1 up to proximal M2) as it is common that LVOs are located there. In Fig.~\ref{fig:markerIdx}, a vessel tree with an LVO on the MCA left branch is shown, visible as abrupt discontinuation in the pathway. Next to the occlusion are the markers, placed where the MCA is expected. In such cases, the nearest points on the vessel mask for each marker coincide on the tip of the interrupted vessel. As a result, the distance with increasing marker index increases linearly, which is shown in Fig.~\ref{fig:markerIdx}. If the slope of the respective linear increase exceeds a certain value (here 2.1), an occlusion in the MCA branch is likely. If no occlusion is present, but due to an unusual pathway deviating from the norm, markers may be misplaced such that their distance exceeds their radii. However, if they are placed in parallel to the existing vessel, their distances to the vessel should be (more or less) constant. Hence their the slope with increasing marker index should be close to 0. This criterion is not suitable for the ICA or PCA, as their trace is too curved to expect apparent linear increases in the distance if the vessel is interrupted.

\section{Results}
To adjust parameters such as thresholds in the pipeline in Fig.~\ref{pipe} and configuration, 28 data sets picked randomly from all data introduced in Sec.~\ref{Data} were used for the development. Hence, 128 subjects are left for final evaluation of the performance.

\subsection{Segmentation and Modelling}
For quantitative evaluation of the segmentation no ground-truth data was available. The annotation of such data is very time consuming but will be part of future work. Consequently the modelling, in particular the DTF-Skeletonization, cannot be quantitatively evaluated as well. Qualitatively, however, small modelling artifacts in forms of artificial loops and bridges may appear in some cases in the surface model if vessel section are very close to or touching each other. Under  \href{https://lme.tf.fau.de/person/thamm/#collapse_1}{this} link we published several videos to demonstrate the interaction with the vessel tree using the vein suppression and the path planning.  

\subsection{Vessel Labeling}
In order to evaluate the labeling on the available data sets, we define the following classes:

Negative class: A vessel belongs to the negative class if (1) it does not exist (2) is completely occluded or (3) is interrupted due to an occlusion. True negatives (TN) are, therefore, vessels labeled as missing (indicated by the removal of a visual marker) because of any of (1), (2), or (3). False negatives (FN) are markers that were not set even though the vessel exists and is neither occluded nor partially occluded, i.e.~none of (1), (2) or (3) are the case. 

Positive class: A vessel exists and is not occluded. A marker placement is true positive (TP) when the marker was correctly placed on a non-occluded existing vessel. False positive (FP) would be the placement of a marker although the vessel fulfills any of the conditions (1), (2), or (3) of the negative class definition. In this error case, a marker is placed on a different vessel nearby for instance. 

The evaluation is particularly strict due to condition (3), as a marker can be set correctly on a more proximal point, e.g.~on the MCA, and still evaluated as FP, since the vessel segment is interrupted by an occlusion distal somewhere else.

\begin{table}
\centering
    \caption{ Performance of the vessel labeling}\label{tab:labeling}
\begin{tabular}[t]{lllllcc}

     Vessel   &   TP &   FN &   TN &   FP  &   Sens. &       Spec. \\
\hline
    ICA L  & 111 & 7 & 1  & 9  & 94.07\% & 10.00\%  \\
    ICA R  & 100 & 9 & 5  & 14 & 91.74\% & 26.32\%  \\
    MCA L  & 91  & 8 & 23 & 6  & 91.92\% & 79.31\%  \\
    MCA R  & 92  & 5 & 17 & 14 & 94.85\% & 54.84\%  \\
    ACA    & 122 & 6 & 0  & 0  & 95.31\% & -- \\
    PCA L  & 118 & 8 & 1  & 1  & 93.65\% & --  \\
    PCA R  & 118 & 9 & 0  & 1  & 92.91\% & --   \\

\end{tabular}
\end{table}%

\begin{table}
\centering
    \caption{Confusion matrix of the LVO patient-wise classification}\label{tab:detection}
    \begin{tabular}[t]{lllc}

     LVO Detection  &   Pred. Pos. &   Pred. Neg &  ~ \\
     \hline
     LVO Pos. & 54 & 27  &  Sens. 66.67\,\% \\
     LVO Neg. & 9  & 38  &  Spec. 80.85\,\%  \\

    \end{tabular}
\end{table}

 Table~\ref{tab:labeling} shows the labeling results. The sensitivity of the labeling ranges from 91\,\% to 95\,\%. As not enough negative cases for the ACA and PCA were available, respective representative specificities were not computable.   
 Tab.~\ref{tab:detection} shows the result of the occlusion detection achieving a sensitivity of 66.67\,\% and specificity of 80.85\,\%. Vessels close to the expected area of an occluded MCA or ICA were often labeled as such, causing true positive labels on occluded branches. This has a consequence on the detection performance and lowers the sensitivity of the LVO detection.  

\section{Discussion}
\label{discussion}

Our vessel path search introduced in Sec.~\ref{SearchOnModels} enabled the interactive vein suppression can provide physicians a tool to better understand the data set at hand in a shorter time by reducing the shown vessels to the relevant ones. As an alternative, similar results can be achieved by the specific acquisition of the arterial phase. However, this cannot be guaranteed, and the success of such an acquisition depends on multiple factors such as the physician's experience or the physiology of the patient. A multi-phase CTA acquisition or a CTP acquisition are alternatives where the arterial phase can be selected retrospectively. Compared to our approach, more than one exposure is required for the multi-phase CTA or CTP. %

The concept of vein suppression could further be extended to detect arteriovenous malformations (AVMs) and AV fistulas by computing the search for two different root nodes, one near the Circle of Willis another one in the Sinus Sagittalis. When the same node shows comparably small distances in the same range to both root nodes, AVMs and AV fistulas might be detected. Diagnosis of such malformations is challenging and they are commonly found incidentally. As no such cases were available in our data, we were not able to put this idea to the test.

As described in Sec.~\ref{SearchOnModels}, further applications could be derived. In this work, we exemplary presented the distance criterion as optimization target. Another optimality criterion would be searching for paths with the largest minimum vessel diameter between two nodes. The diameter optimization is a min-max problem, as the minimum diameter that occurs in a vessel path is the one that needs to be maximized.
Based on the first node property, the path search could guide interventional treatment, where paths need to be chosen that are large enough for a particular catheter to fit through. Using the second property, vessel sections which are not passable for a given catheter could be concealed, such that physicians can easily exclude unsuitable pathways as they perform the intervention. Rapid changes in the edge direction, which are available due to the third property, might be indicators for anatomical abnormalities. For instance, if two connected edges point into two opposite directions, a stenosis can probably be located close to the respective node.

Related work by \cite{Suran.} or ~\cite{shen2020automatic} used the Dijkstra algorithm for the distance optimized path search, whereas we chose A* combined with the Euclidean heuristic. A* is optimally efficient for any given consistent admissible heuristic, which means no other optimal algorithm that utilizes the same heuristic expands fewer nodes than A* (except possible tie-breaks) \citep{StuartJ.RussellandPeterNorvig.2010}. Both algorithms lead to the same results, but A* turns out to be faster in practical problems. Comparing Dijkstra and A* based on their time complexity is difficult, as the exact time complexity for an algorithm enhanced with an admissible heuristic is rather complex and hard to estimate \citep{RichardE.KorfandMichaelReid.}. For a graph $\mathcal{G} = \{\mathcal{V}, \mathcal{E}\}$ with nodes $\mathcal{V}$, edges $\mathcal{E}$, branching factor of $b$ and the depth of the optimal solution of $d$ the Dijkstra algorithm shows in the worst case a linearithmic complexity $\mathcal{O}(|\mathcal{V}|\log(|\mathcal{V}|) + |\mathcal{E}|)$ with $|\mathcal{V}|$ and $|\mathcal{E}|$ as the number of nodes and edges \citep{cormen2009introduction}. In contrast, the A* algorithm can lead in the worst case to an exponential time complexity $\mathcal{O}(b^d)$ \citep{StuartJ.RussellandPeterNorvig.2010}, as potentially a bad heuristic can guide the search to all other nodes first until the algorithm ends up in the goal node. Thus, the actual time complexity depends firstly on the problem at hand and secondly on the choice of heuristic and how it explores new sub-branches, the resulting branching factors, and the final depth from the start node to the goal node \citep{StuartJ.RussellandPeterNorvig.2010}. Nevertheless, it is generally recommended to prefer a heuristic search, so we compared the computation time empirically instead. The graph search described in \ref{SearchOnModels} on a data set with a large injected contrast bolus based on the same root node and averaged over 100 repetitions, took 1088\,ms per run with the Dijkstra algorithm and 723\,ms per run with the A* algorithm using the Euclidean heuristic. Hence, the use of A* decreased the run time to 64\,\%. Furthermore, \cite{Suran.} applied their algorithm on a pixel basis. We instead applied the algorithm on the bifurcation level. The bifurcation search considerably increases the performance because in \cite{Suran.} each pixel represents one node, and as shown before, the time complexity grows linearithmically with the number of nodes. Combining the A* search, and the reduction to a search on bifurcation level allows more interactive use of our approach without any high-end hardware. In \cite{Suran.} no considerations were made about the case when vessel trees are not connected to the main tree. As this is crucial, especially if occlusions separate two vessel trees, we have extended our algorithm accordingly (see Sec. \ref{SearchOnModels}).

Since the labeling task is combined with the occlusion detection task, a fair comparison to existing works is challenging, as other works do not address both problems at once. However, several steps can further be optimized in order to increase especially the sensitivity of the detection. For instance, the registration should be done using different atlas maps, e.g.~one representing the skull, as bone structures appear more dominant in CT images as brain tissue. More rules to evaluate missing markers, and hence occluded vessel traces should be explored as well. A semi-automated process to evaluate the labeling results would further ease the development of such rules. 

\section{Conclusions}
The work at hand has shown a pipeline that segments the intracranial vessel tree and subsequently constructs a geometric, graphical model of it. We have shown that the systematic application of a path search algorithm on all nodes of the tree model leads to various applications, dependent on the chosen optimality criterion for the path search. Exemplary we have chosen the shortest pathway search as optimization criterion, from which we were able to show two applications, namely the path search itself, which supports the interactive planning for interventional treatment, and the interactive suppression of veins to reduce the vessel tree to parts which are of interest in the diagnosis of strokes.

We further extended the pipeline with the automated labeling of the cerebrovascular system, namely MCA, ICA, PCA, each left and right, and ACA. On 128 patients we have achieved labeling sensitivities between 91\,\% and 95\,\%. As two third of the patients we evaluated in this study suffered from LVOs, we extended the labeling algorithm to detect LVOs as well. In this regard, we have achieved a sensitivity of 67\,\% and a specificity of 81\,\%. Although the labeling itself works reliably and is comparable to state of the art in MRI, we consider the occlusion detection rather as a proof of concept and suggest further investigation of the potential of this method.

\bibliographystyle{unsrtnat}
\bibliography{references}  %

\end{document}